# Ideal Nodal Rings of One-Dimensional Photonic Crystals in the Visible Region

Wei-Min Deng, Ze-Min Chen, Meng-Yu Li, Chao-Heng Guo, Xiao-Dong Chen, Wen-Jie Chen[*], and Jian-Wen Dong[*]

*School of Physics & State Key Laboratory of Optoelectronic Materials and Technologies, Sun Yat-sen University, Guangzhou 510275, China.*

[*]Corresponding author: chenwenj5@mail.sysu.edu.cn; dongjwen@mail.sysu.edu.cn

**Abstract**

Three-dimensional (3D) artificial metacrystals host rich topological phases, such as Weyl points, nodal rings and 3D photonic topological insulators. These topological states enable a wide range of applications, including 3D robust waveguide, one-way fiber and negative refraction of surface wave. However, these carefully designed metacrystals are usually very complex, hindering their extension to nanoscale photonic systems. Here, we theoretically proposed and experimentally realized an ideal nodal ring in visible region using a simple 1D photonic crystal. The π-Berry phase around the ring is manifested by a 2π reflection phase's winding and the resultant drumhead surface states. By breaking the inversion symmetry, the nodal ring can be gapped and the π-Berry phase would diffuse into a toroidal shaped Berry flux, resulting in photonic ridge states (the 3D extension of quantum valley Hall states). Our results provide a simple and feasible platform for exploring 3D topological physics and their potential applications in nanophotonics.

**Introduction**

In the study of topological states, topological semimetals[1-3], featured by their symmetry-protected band degeneracies, serve as the parent states of various types of topological gapped states and have generated much research interest. These gapless band structures were usually studied in two- or three-dimensional (3D) lattice crystal, good examples are graphene[4,5] and Dirac/Weyl semimetal[6,7]. Depending on the symmetries of the system, these stable degeneracies can occur at isolated points[6,7], along closed lines[8,9] or even on surfaces[10,11] in 3D momentum space. In principle, for a system with higher symmetry, the topological band touching would occur in higher dimensional manifold. By lowering the crystal symmetries through certain types of interaction (spin-orbit interaction or external field), these degeneracies can be lifted, and the gapless states would transit to a rich variety of gapped topological states (such as topological insulating states or 3D quantum Hall states), along with the in-gap excitations guaranteed by the nontrivial band topology.

Apart from topological states of matter, topological band theory and the relevant concepts apply equally well to the photonic system[12-15], and have recently inspired many novel applications in nanophotonics, such as backscattering-immune waveguides[16-20], robust delay lines[21] and high-performance lasers[22-25]. Most of these devices were based on 2D lattice crystals (e.g. photonic quantum spin/valley Hall systems) for their easier fabrication. But the 3D topological states and their topological effects in optical wavelength scale remain untamed, mostly because they are usually accompanied with complex 3D structures[26-33]. This may limit the development of topological photonics and its potential application in nanophotonics. Thus, it would be highly desirable if one could achieve 3D topological state using a simple structure, preferably a 1D crystal.

Here, we theoretically propose and experimentally realize an ideal nodal ring state and the relevant topological gapped states in simple 1D photonic crystals (PCs). By taking off-axis momenta into account, we find that even a periodic layered medium can exhibit an ideal Dirac nodal ring dispersion without frequency variation in its 3D momentum space. By lowering the crystal's symmetry, the ring degeneracies can transit to a rich variety of topological gapped states, such as photonic ridge state, the 3D extension of quantum valley Hall effect. These gapless/gapped bulk bands and the topologically-protected surface states are experimentally observed by angle-resolved reflection spectra in the visible region. Their nontrivial band topologies are analyzed through effective Hamiltonian and the calculated Berry curvature. Our results demonstrate a feasible platform for studying the 3D topological phases of light in nanoscale and exploring their potential applications in nanophotonics.

**Results**

Nodal ring can be deemed as the extrusion of 2D Dirac points along a closed loop and thus it carries a quantized Berry phase of π. Its low energy Hamiltonian takes the form of $H_N = \mu(k_r^2 - k_0^2)\sigma_z + v_z k_z \sigma_x + v k_r^2 \sigma_0$, where $k_0$ denotes the radius of nodal ring and $\sigma_i$ is Pauli matrix. On each cut plane containing $k_z$ (e.g. the $k_y$-$k_z$ plane), the Hamiltonian reduces to a 2D Dirac Hamiltonian. According to the band tilting condition (the coefficients $v$ and $\mu$), nodal rings can be classified into type I/II, as sketched in Figs.1a & 1b. When $|v/\mu| > 1$, the slopes of the two bands have same signs, resulting in the tilted crossings along the nodal ring. Note that this tilting would not affect the topological characteristic of nodal ring. These nodal line degeneracies, either type-I or type-II, are usually protected by PT symmetry. By breaking P or T symmetry (introducing a mass term $m_y(\mathbf{k})\sigma_y$ in $H_N$), the nodal ring can break into multiple nodal points or be totally gapped, which depends on the specific form of the mass term $m_y(\mathbf{k})$. For the simplest case that $m_y$ is a constant, the nodal ring would be totally gapped and has a ridge-like dispersion, as sketched in Figs. 1c & 1d. These ridge states can be deemed as the 3D version of quantum valley Hall effect.

In addition, these simplified models are isotropic along any radial direction, which is not the case for a real crystal. Even for a 3D crystal with discrete rotational symmetry, the nodal crossings are not guaranteed to occur at the same radius or the same eigen frequency/energy. Ideal nodal ring can in principal exist in the systems with continuous rotational symmetry, for example, a layered medium or 1D PC. For a 1D layered medium, apart from the Bloch $k_z$ in the out-of-plane direction, we take the in-plane components $k_x$ and $k_y$ into account. Then we can have a 3D momentum space for a 1D lattice structure and explore its 3D band topology.

Consider a typical 1D PC with two dielectric components, as shown in Fig. 2a. Since the crystal has discrete translational symmetry in *z*-direction but continuous translational symmetry in *xy*-plane, the Brillouin zone of 1D PC is an infinite slab bounded by two planes of $k_z = \pm \pi/a$ (cyan planes in Fig. 2b). The left panel of Fig. 2c plots the $k_z$-dispersion for normal propagation. A band gap from 0.27c/a to 0.33c/a lies between the lowest two bands, which are doubly-degenerate for s and p polarizations. As the in-plane component $k_y$ increases (the right panel of Fig. 2c), the four bands split and curve upward with different velocities. The two p-polarized bands (blue and violet) linearly cross at $k_y = 0.513 \times 2\pi/a$. In fact, this is a type-II Dirac cone on the $k_y$-$k_z$ plane as illustrated by Fig. 2d. Since 1D PC is rotation-invariant, this tilted cone extrudes along azimuthal direction (Fig. 2e), forming an ideal Dirac nodal ring without frequency variation.

To confirm the band topology of the nodal ring, we derive its $\mathbf{k} \cdot \mathbf{p}$ Hamiltonian via transfer matrix method and it reads

$$H = v_x \xi_{k_z} \sigma_x + (v_0 \sigma_0 + v_z \sigma_z) \xi_{k_r} \tag{1}$$

where $\xi_{k_z} = (k_z - k_{z0})/k_{z0}$, $\xi_{k_r} = (k_r - k_{r_0})/k_{r_0}$. $k_{z0}$ denotes the $k_z$-plane the nodal ring lies on, $k_{r_0}$ is the radius of nodal ring (see more details in Supplementary Information). On any $k_r$-$k_z$ cut plane, the effective Hamiltonian resembles a type-II Dirac Hamiltonian, implying the $\pi$ Berry phase it carries[34]. Interestingly, the Dirac crossing in Fig. 2c happens to occur at the Brewster angle between the two dielectric materials[35]. But the Brewster effect is not a necessary condition for the existence of nodal ring. Even for the 1D PCs without Brewster effect, the nodal ring still exists as long as the inversion symmetry is preserved (see Supplementary Fig. 2).

For its simple 1D structure, such kind of nodal ring photonic crystals (NRPCs) can be readily fabricated using current nanofabrication techniques. Here, silicon nitride and silica are chosen to demonstrate our idea in the visible region. Figure 3a shows the scanning-electron-microscope (SEM) image of an 8-period NRPC fabricated using Chemical Vapor Deposition. In our experiment, the nodal ring states are investigated through angle-resolved reflection measurement, using the configuration depicted in Fig. 3b. Since the Dirac node lies below the light cone of air (Fig. 2c), the photonic states near the nodal ring cannot be directly excited by an incident beam from air. Hence, a truncated hemi-cylindrical prism (JGS2 quartz glass, $n_p = 1.45$) is put on the upper sample to couple the incident beam into the NRPC. Figure 3c shows the calculated reflection spectra for p-polarization as a function of $k_y = n_p k_0 \sin\theta_i$, where $\theta_i$ is the incident angle. As predicted by the bulk dispersion in Fig. 2c, the band gap closes and reopens at $k_y = 0.513 \times 2\pi/a$, corresponding to the position of nodal point. Figure 3d shows the virtual 3D reflection spectra based on our measured reflection data along $k_y$ direction. One can see that there are two high-reflection regions, corresponding to two band gaps connected by the nodal point. In contrast, the reflection is very low near the nodal point, due to the excitation of the nodal ring states.

The topological features of nodal rings are manifested in the surface properties when the NRPC is truncated in one dimension. Due to the nontrivial Berry phase around the nodal ring, the surface momentum space is divided into two regions (inside or outside the ring) with $\pi/0$ Zak phases, indicating the existence of a drumhead surface state pinned at the nodal ring. Experimentally, a 40nm-thicked silver film is deposited on the NRPC with 8 periods to study its surface, as shown in Fig. 4a. Figure 4c plots the surface dispersion between NRPC and silver, where a p-polarized surface band (red line) is pinned at the nodal point (gray circle) and exhibits a drumhead dispersion inside the nodal ring. Besides, an s-polarized surface band exists and is degenerate with the p-polarized band at $k_r = 0$ (for normal incidence). These surface bands can be clearly seen in our measured angle-resolved reflection data in Fig. 4d. In fact, the existence of this drumhead surface state is ubiquitous and does depend on the surface truncation or the other gapped material (see more results in Supplementary Information). To gain a deep insight into the underlying mechanism, we calculate the reflection phase of NRPC in Fig. 4b. The reflection phase $\varphi_{PC}$ exhibits a $2\pi$-winding near the nodal point and the nodal point serves as a singularity of reflection phase. Since a stable surface state should satisfy the condition of $\varphi_{PC} + \varphi_{Ag} = 2N\pi$, there would always be a surface band pinned at the nodal point, no mater the reflection properties of the other material.

In 2D photonic system, Dirac cones can be gapped by inversion symmetry breaking and results in the photonic valley Hall states[36-40]. As its 3D extrusion along the azimuthal angle, the nodal ring degeneracy can be lifted and deform into a ridge-like band structure (Fig. 1d), here we term the crystal with ridge dispersion as ridge photonic crystal (RPC). As shown in Fig. 5a, we break the inversion symmetry by replacing the bottom (top) layer A by a layer C with refractive index of 3, named as $RPC_1$ ($RPC_2$). Figure 5b shows the bulk band structure of $RPC_1$. One can see that the original nodal ring degeneracy is lifted, leading to a partial band gap. Its effective Hamiltonian is written as $H_1 = \mu(k_r^2 - k_0^2)\sigma_z + v_z k_z \sigma_x + v k_r^2 \sigma_0 + m\sigma_y$, where the symmetry breaking introduces a rotational-invariant mass term into $H_1$. Meanwhile, the π Berry phase localized at the nodal ring would spread out into a toroidal shaped Berry flux flowing in the counter-clockwise direction. Figure 5c show the lower band's Berry curvature of $RPC_1$, whose inset plots a zoom-in view on the cut plane of $k_y = 0$. Since $RPC_1$ and $RPC_2$ are inversion partners of each other, they share an identical bulk band dispersion but have different eigen wave functions with opposite Berry curvatures. Numerical calculations in Figs. 5c & 5d show that the Berry vortexes of $RPC_1$ and $RPC_2$ are localized near the ridges and flow in opposite direction, indicating their distinct topological properties.

When two RPCs with distinct topological properties are joined together and form a domain wall, an in-gap surface state traversing the band gap is expected. Figures 5e-5h discuss two surface configurations, $RPC_1$-$RPC_2$ domain wall (Fig. 5e) and $RPC_2$-$RPC_1$ domain wall (Fig. 5f). Both calculated surface dispersions in Figs. 5e & 5f have a gapless p-polarized surface band (red line), which implies a conical-frustum shaped dispersion near the ridge. It means that these topological surface states can propagate in all directions along the interface between two RPCs. In experiment, we fabricate two domain wall samples with different configurations. The left panel in Fig. 5a shows the SEM image of the $RPC_2$-$RPC_1$ domain wall. The measured angle-resolved spectra in Figs.5g & 5h show good agreement with the calculated dispersions and demonstrate the gapless surface states between two RPCs.

**Conclusion**

In conclusion, we propose and experimentally demonstrate an ideal type-II nodal ring using simple 1D structures. As the most remarkable feature of Dirac nodal ring, the drumhead surface states pinned at the ring are experimentally observed. By breaking inversion symmetry, the nodal ring state can transit to a photonic ridge state, whose nontrivial topology is signified by a clockwise/counter-clockwise Berry vortex in momentum space. The topological surface states between two RPCs with opposite Berry vortexes are also demonstrated. Compared to the complex 3D metamaterials, our 1D PCs are easier to design and fabricate, which facilitate the future application of 3D topological states in nanophotonics, such as resonant scattering[41] and negative refraction[42]. Besides, since the nodal ring can transform to Weyl point by symmetry breaking[43], our work make it possible to realize Weyl point and associated topological phenomena in 1D PCs.

## Methods

**Simulations.** The band structures were calculated by MIT Photonic Bands (MPB)[44], which is based on the plane-wave expansion (PWE) method. Reflection spectra were calculated by the transfer matrix method.

**Sample fabrication.** Films are deposited on a quartz substrate by inductance coupled plasma enhanced chemical vapor deposition system (ICP-CVD, Oxford Instruments PlasmaPro System 100). The deposition rate is 12 nm/min for $SiO_2$, 13.5 nm/min for $Si_xN_y$ and 7 nm/min for silicon-rich nitride. The Ag film is deposited by electron beam evaporation (DE400DHL, DE Technology Inc.) and the corresponding deposition rate is 6 nm/min.

**Optical measurements.** The refractive index and extinction coefficient of silicon-rich nitride were determined from spectrometry ellipsometry measurements (Sentech SE400). Angle-resolved reflection spectra of films were measured by an angle-resolved spectrum system (R1-UV, Ideaoptics, PR China). In order to avoid the influence of reflected light from the interface between the substrate and air, the bottom surface of the substrate is contacted to an extra thick glass substrate (25mm*25mm*20mm) by means of index-matching liquid.

**Acknowledgements**

We thank Prof. Shaoji Jiang for the help in sample preparation. This work was supported by National Natural Science Foundation of China (Grant Nos. 12074443, 11874435, 11904421, and 62035016), the Young Top-Notch Talent for Ten Thousand Talent Program (2020-2023), Guangdong Basic and Applied Basic Research Foundation (Grant No. 2019B151502036), Natural Science Foundation of Guangdong Province (Grant No. 2018B030308005), Guangzhou Science, Technology and Innovation Commission (Grant Nos. 201904010223, and 202102020693), Fundamental Research Funds for the Central Universities (Grant No. 20lgzd29, 20lgjc05 and 2021qntd27).

**References**

1  Fang, C., Weng, H., Dai, X. & Fang, Z. Topological nodal line semimetals. *Chin. Phys. B* **25**, 117106 (2016).
2  Hirayama, M., Okugawa, R. & Murakami, S. Topological Semimetals Studied by Ab Initio Calculations. *J. Phys. Soc. Jpn.* **87**, 041002 (2018).
3  Armitage, N. P., Mele, E. J. & Vishwanath, A. Weyl and Dirac semimetals in three-dimensional solids. *Rev. Mod. Phys.* **90**, 015001 (2018).
4  Novoselov, K. S. et al. Two-dimensional gas of massless Dirac fermions in graphene. *Nature* **438**, 197-200 (2005).
5  Zhang, Y. B., Tan, Y. W., Stormer, H. L. & Kim, P. Experimental observation of the quantum Hall effect and Berry's phase in graphene. *Nature* **438**, 201-204 (2005).
6  Young, S. M. et al. Dirac Semimetal in Three Dimensions. *Phys. Rev. Lett.* **108**, 140405 (2012).
7  Wan, X., Turner, A. M., Vishwanath, A. & Savrasov, S. Y. Topological semimetal and Fermi-arc surface states in the electronic structure of pyrochlore iridates. *Phys. Rev. B* **83**, 205101 (2011).
8  Fang, C., Chen, Y., Kee, H.-Y. & Fu, L. Topological nodal line semimetals with and without spin-orbital coupling. *Phys. Rev. B* **92**, 081201(R) (2015).
9  Bzdušek, T., Wu, Q., Rüegg, A., Sigrist, M. & Soluyanov, A. A. Nodal-chain metals. *Nature* **538**, 75-78 (2016).
10 Liang, Q.-F., Zhou, J., Yu, R., Wang, Z. & Weng, H. Node-surface and node-line fermions from nonsymmorphic lattice symmetries. *Phys. Rev. B* **93**, 085427 (2016).
11 Fu, B. B. et al. Dirac nodal surfaces and nodal lines in ZrSiS. *Sci. Adv.* **5**, eaau6459 (2019).
12 Lu, L., Joannopoulos, J. D. & Soljačić, M. Topological photonics. *Nat. Photon.*

**8**, 821-829 (2014).

13  Ozawa, T. et al. Topological photonics. *Rev. Mod. Phys.* **91**, 015006-015081 (2019).

14  Kim, M., Jacob, Z. & Rho, J. Recent advances in 2D, 3D and higher-order topological photonics. *Light Sci. Appl.* **9**, 130-159 (2020).

15  Yuan, L. Q., Lin, Q., Xiao, M. & Fan, S. H. Synthetic dimension in photonics. *Optica* **5**, 1396-1405 (2018).

16  Shalaev, M. I., Walasik, W., Tsukernik, A., Xu, Y. & Litchinitser, N. M. Robust topologically protected transport in photonic crystals at telecommunication wavelengths. *Nat. Nanotechnol.* **14**, 31-34 (2018).

17  He, X.-T. et al. A silicon-on-insulator slab for topological valley transport. *Nat. Commun.* **10**, 872-880 (2019).

18  Parappurath, N., Alpeggiani, F., Kuipers, L. & Verhagen, E. Direct observation of topological edge states in silicon photonic crystals: Spin, dispersion, and chiral routing. *Sci. Adv.* **6**, eaaw4137 (2020).

19  Barik, S. et al. A topological quantum optics interface. *Science* **359**, 666-668 (2018).

20  Ma, J., Xi, X. & Sun, X. Topological Photonic Integrated Circuits Based on Valley Kink States. *Laser Photon. Rev.*, 1900087 (2019).

21  Hafezi, M., Demler, E. A., Lukin, M. D. & Taylor, J. M. Robust optical delay lines with topological protection. *Nat. Phys.* **7**, 907-912 (2011).

22  Bandres, M. A. et al. Topological insulator laser: Experiments. *Science* **359**, eaar4005 (2018).

23  Bahari, B. et al. Nonreciprocal lasing in topological cavities of arbitrary geometries. *Science* **358**, 636-640 (2017).

24  Yang, Z. Q., Shao, Z. K., Chen, H. Z., Mao, X. R. & Ma, R. M. Spin-Momentum-Locked Edge Mode for Topological Vortex Lasing. *Phys. Rev. Lett.* **125**, 013903 (2020).

25  Zeng, Y. et al. Electrically pumped topological laser with valley edge modes. *Nature* **578**, 246-250 (2020).

26  Lu, L. et al. Symmetry-protected topological photonic crystal in three dimensions. *Nat. Phys.* **12**, 337-340 (2016).

27  Slobozhanyuk, A. et al. Three-dimensional all-dielectric photonic topological insulator. *Nat. Photon.* **11**, 130-136 (2017).

28  Yang, Y. et al. Realization of a three-dimensional photonic topological insulator. *Nature* **565**, 622–626 (2019).

29  Lu, L. et al. Experimental observation of Weyl points. *Science* **349**, 622-624 (2015).

30  Chen, W. J., Xiao, M. & Chan, C. T. Photonic crystals possessing multiple Weyl points and the experimental observation of robust surface states. *Nat. Commun.* **7**, 13038 (2016).

31  Noh, J. et al. Experimental observation of optical Weyl points and Fermi arc-like surface states. *Nat. Phys.* **13**, 611-617 (2017).

32  Gao, W. et al. Experimental observation of photonic nodal line degeneracies in

| | |
|---|---|
| | metacrystals. *Nat. Commun.* **9**, 950 (2018). |
| 33 | Yang, B. et al. Ideal Weyl points and helicoid surface states in artificial photonic crystal structures. *Science* **359**, 1013-1016 (2018). |
| 34 | Li, S. et al. Type-II nodal loops: Theory and material realization. *Phys. Rev. B* **96**, 081106(R) (2017). |
| 35 | Fink, Y. et al. A Dielectric Omnidirectional Reflector. *Science* **282**, 1679 (1998). |
| 36 | Ma, T. & Shvets, G. All-Si valley-Hall photonic topological insulator. *New J. Phys.* **18**, 025012-025020 (2016). |
| 37 | Chen, X.-D., Zhao, F.-L., Chen, M. & Dong, J.-W. Valley-contrasting physics in all-dielectric photonic crystals: Orbital angular momentum and topological propagation. *Phys. Rev. B* **96**, 020202(R)-020206(R) (2017). |
| 38 | Dong, J.-W., Chen, X.-D., Zhu, H., Wang, Y. & Zhang, X. Valley photonic crystals for control of spin and topology. *Nat. Mater.* **16**, 298-302 (2017). |
| 39 | Gao, F. et al. Topologically protected refraction of robust kink states in valley photonic crystals. *Nat. Phys.* **14**, 140-145 (2017). |
| 40 | Wu, X. et al. Direct observation of valley-polarized topological edge states in designer surface plasmon crystals. *Nat. Commun.* **8**, 1304-1312 (2017). |
| 41 | Zhou, M. et al. Electromagnetic scattering laws in Weyl systems. *Nat. Commun.* **8**, 1388 (2017). |
| 42 | Yang, B. et al. Momentum space toroidal moment in a photonic metamaterial. *Nat. Commun.* **12**, 1784 (2021). |
| 43 | Yan, B. & Felser, C. Topological Materials: Weyl Semimetals. *Annual Review of Condensed Matter Physics* **8**, 337-354 (2017). |
| 44 | Johnson, S. G. & Joannopoulos, J. D. Block-iterative frequency-domain methods for Maxwell's equations in a planewave basis. *Opt. Express* **8**, 173-190 (2001). |

**Figures and Figure Captions**

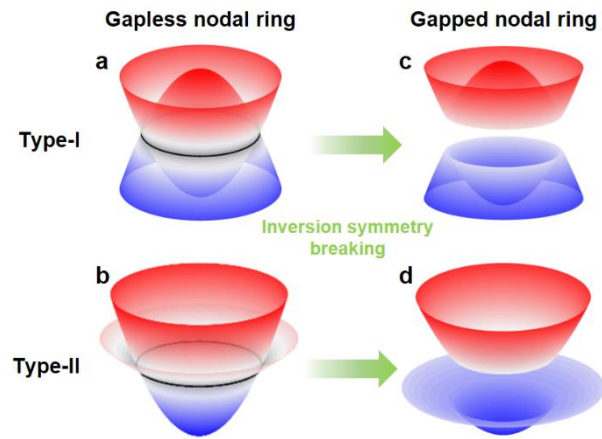

Fig. 1. Schematics of type-I and type-II Dirac nodal rings. a-b, gapless nodal rings protected by inversion symmetry. c-d, gapped nodal rings after inversion symmetry breaking which exhibit ridge-like dispersion.

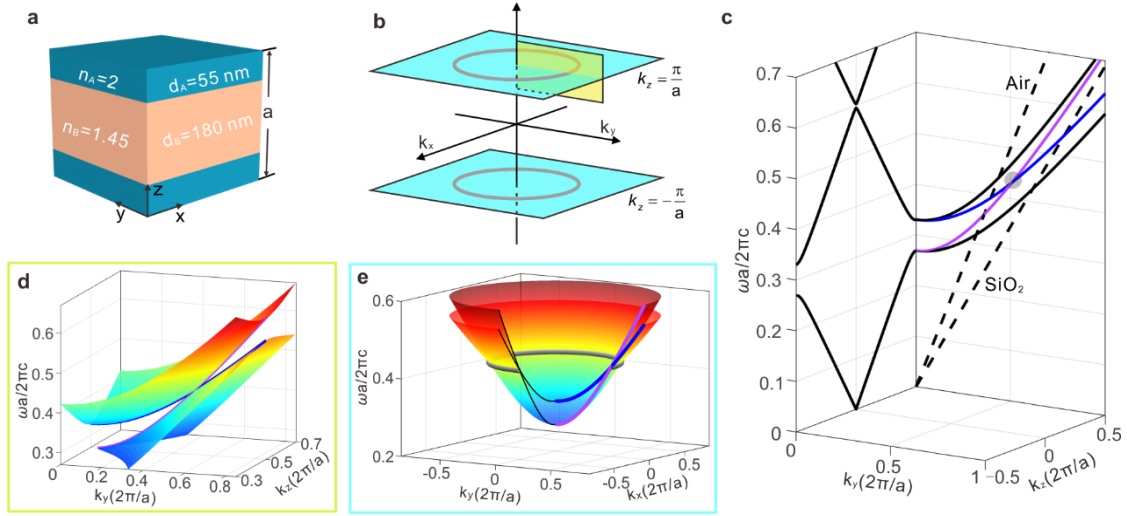

Fig. 2. One-dimensional (1D) photonic crystal with ideal Type-II nodal ring. (a) Unit cell of the 1D nodal ring photonic crystal (NRPC). It consists of three dielectric layers (A/B/A) and obeys inversion symmetry. The refractive index and thickness of layer A (B) is 2 (1.45) and 55 nm (180 nm), respectively. (b) 3D Brillouin zone of the 1D crystal. The two cyan planes highlight the Brillouin zone boundaries at $k_z = \pm \pi/a$, on which the gray circles mark the nodal ring. (c) Band dispersions along $k_z$ and $k_y$ directions. The dashed lines represent the light lines of air and SiO$_2$. (d) Zoom in view of the tilted Dirac cone on the $k_y - k_z$ plane highlighted by yellow in (b). (e) Band dispersion on the $k_x - k_y$ plane with $k_z = \pi/a$, which has a type-II nodal ring crossing (gray torus). For a better view of the nodal ring, only three quarters of the bands are shown.

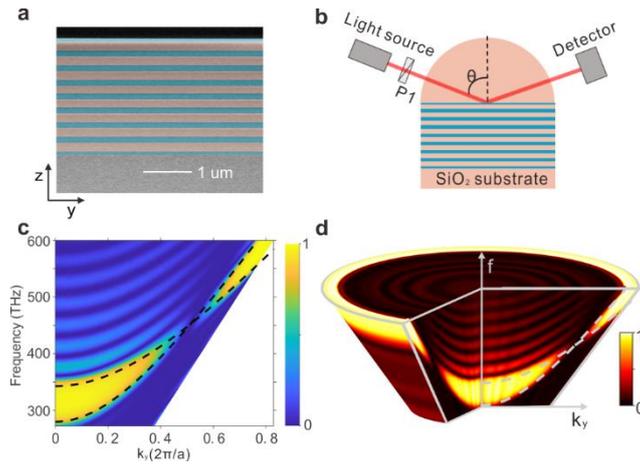

Fig. 3. Experimental demonstration of the type-II nodal ring. (a) Scanning-electron-microscope (SEM) image of the fabricated sample. (b) Experimental setup for measuring the angle-resolved reflection spectra. The sample is contacted to a half cylindrical prism. P1, polarizer. (c/d): Measured/calculated angle-resolved reflection spectra for p-polarized incident light. The dashed curves represent the gap edges of the NRPC.

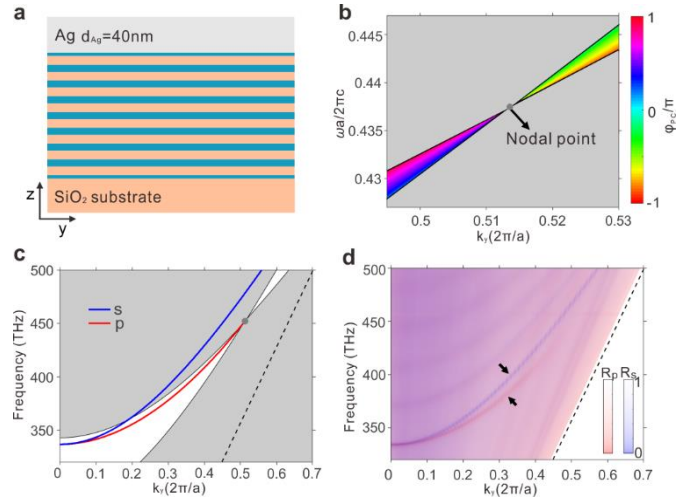

Fig. 4. Drumhead surface states of the 1D gapless NRPC. (a) Schematic of the interface between 1D NRPC and silver. The silver film with a thickness of 40 nm is deposited on the 1D NRPC with 8 periods. (b) Calculated p-polarized reflection phase ($\varphi_{PC}$) of the NRPC. Note that in the vicinity of the Dirac nodal ring, there is a $2\pi$ phase winding around the nodal point which manifests its nontrivial Berry phase. (c) Projected band structure with s- (blue line) and p- (red line) polarized surface states. P-polarized surface states are pinned at the nodal point due to the phase winding around the nodal point. The black dashed line represents the light line of $SiO_2$. (d) Overlaid plot of the measured angle-resolved reflection spectra for both polarizations. Two black arrows highlights two lines of reflection dips corresponding to the surface states.

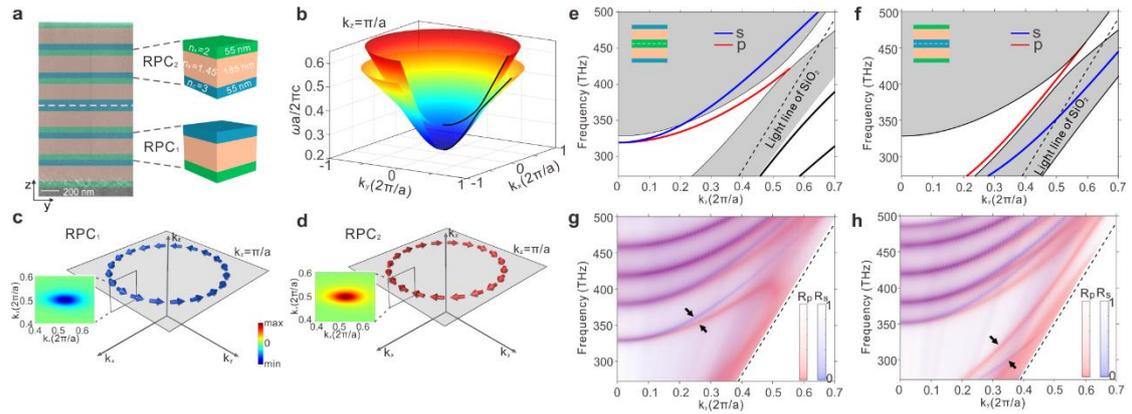

Fig. 5. Gapped ridge state induced by inversion symmetry breaking. (a) Left panel: SEM image of the RPC$_2$-RPC$_1$ domain wall. The white dashed line highlights the boundary. Right panel: Two types of ridge photonic crystals (RPC$_1$ and RPC$_2$). (b) Bulk band dispersion of RPC$_1$ on the plane of $k_z = \pi/a$ with a partial gap opened in the vicinity of nodal ring. (c) & (d): Localized Berry vortexes for two RPCs. The insets show the cross-sectional views of the Berry curvature. Note that the Berry vortexes of two RPCs flow in opposite directions (clockwise or counter-clockwise) since they are linked by inversion. (e) Projected band of RPC$_1$-RPC$_2$ domain wall. Gapless surface states traverse the partial gap due to the opposite Berry vortexes of the two RPCs. (g) Measured angle-resolved reflection spectra for both polarizations. Two lines of reflection dips (highlighted by the black arrows) inside the band gap demonstrates the topologically protected surface states. (f) & (h): Similar to (e) & (g), but for RPC$_2$-RPC$_1$ domain wall.